\documentstyle [12pt,eqsecnum,aps,amsfonts] {revtex}
\input epsf
\newcommand{\beq}{\begin{equation}}
\newcommand{\eeq}{\end{equation}}
\newcommand{\beqs}{\begin{eqnarray}}
\newcommand{\eeqs}{\end{eqnarray}}

\begin{document}
\draft

\baselineskip 6.0mm

\title{Ground State Entropy of Potts Antiferromagnets and the 
Approach to the 2D Thermodynamic Limit} 

\author{Robert Shrock\thanks{email: shrock@insti.physics.sunysb.edu}
\and Shan-Ho Tsai\thanks{email: tsai@insti.physics.sunysb.edu}}

\address{
Institute for Theoretical Physics  \\
State University of New York       \\
Stony Brook, N. Y. 11794-3840}

\maketitle

\vspace{10mm}

\begin{abstract}

   We study the ground state degeneracy per site (exponent of the ground state
entropy) $W(\Lambda,(L_x=\infty) \times L_y,q)$ for the $q$-state Potts
antiferromagnet on infinitely long strips with width $L_y$ of 2D lattices
$\Lambda$ with free and periodic boundary conditions in the $y$ direction,
denoted FBC$_y$ and PBC$_y$.  We show that the approach of $W$ to its 2D
thermodynamic limit as $L_y$ increases is quite rapid; for moderate values of
$q$ and $L_y \simeq 4$, $W(\Lambda,(L_x=\infty) \times L_y,q)$ is within about
5 \% and ${\cal O}(10^{-3})$ of the 2D value $W(\Lambda,(L_x=\infty) \times
(L_y=\infty),q)$ for FBC$_y$ and PBC$_y$, respectively.  The approach of $W$ to
the 2D thermodynamic limit is proved to be monotonic (non-monotonic) for
FBC$_y$ (PBC$_y$).  It is noted that ground state entropy determinations
on infinite strips can be used to obtain the central charge for cases with
critical ground states.

\end{abstract}


\vspace{16mm}

\pagestyle{empty}
\newpage

\pagestyle{plain}
\pagenumbering{arabic}
\renewcommand{\thefootnote}{\arabic{footnote}}
\setcounter{footnote}{0}

\section{Introduction}

    Nonzero ground state entropy, $S_0 \ne 0$, is an important subject in
statistical mechanics.  One physical example is provided by ice, for which the
residual molar entropy is $S_0 = 0.82 \pm 0.05$ cal/(K-mole), i.e., $S_0/R =
0.41 \pm 0.03$, where $R=N_{Avog.}k_B$ \cite{lp,liebwu}.  This is equivalent to
a ground state degeneracy per site $W > 1$, since $S_0 = k_B \ln W$.  Such
nonzero ground state entropy violates the third law of thermodynamics (see,
e.g., \cite{al,chowwu}). The $q$-state Potts antiferromagnet (AF)
\cite{potts,wurev} exhibits nonzero ground state entropy (without frustration)
for sufficiently large $q$ on a given lattice $\Lambda$ and serves as a useful
model for the study of this phenomenon.  There is an interesting connection
with graph theory here, since the zero-temperature partition function of the
above-mentioned $q$-state Potts antiferromagnet on a graph $G$ satisfies
$Z(G,q,T=0)_{PAF}=P(G,q)$, where $P(G,q)$ is the chromatic polynomial
expressing the number of ways of coloring the vertices of the graph $G$ with
$q$ colors such that no two adjacent vertices have the same color \cite{rtrev}.
Thus, \beq W([\lim_{n \to \infty} G \ ],q) = \lim_{n \to \infty} P(G,q)^{1/n}
\label{w}
\eeq where $n=v(G)$ is the number of vertices of $G$.  Nontrivial exact
solutions for this function are known in only a very few cases, all for 2D
lattices: the square lattice for $q=3$ \cite{lieb}, triangular lattice and, for
$q=3$, the kagom\'e lattice \cite{baxter}.  Of course, one can use large--$q$
series expansions \cite{series,kewser,ww,w3,wn}, rigorous upper and lower
bounds \cite{biggs,ww,w3,wn}, and Monte Carlo simulations (see, e.g.,
\cite{cp,wsk,p3afhc}).  It is also worthwhile to generalize $q$ from ${\mathbb
Z}_+$ to ${\mathbb C}$ and study $W(\{G\},q)$ in the complex $q$ plane for
infinite-$n$ limits of various families of graphs, $\{G\}$
\cite{bds}-\cite{hs}.  On the positive real axis, $W(\{G\},q)$ is an analytic
function down to a point which we denote $q_c(\{G\})$ \cite{w}. 

Since it is possible to obtain exact analytic solutions for infinitely long
strips of 2D lattices \cite{strip}, one has an alternate way to investigate
$W(\Lambda,q)$ for 2D lattices, namely to calculate exactly $W$ on such
infinitely long strips of progressively greater widths.  These $W$ functions
for infinitely long strips have interesting analytic structure in their own
right, which was investigated in detail in Ref. \cite{strip}.  Here we shall
use them for a different purpose: to investigate how rapidly the 2D
thermodynamic limit is approached as the width of the strips increases.  We
find that this approach is quite rapid.  Of course, to demonstrate this does
not require the use of exact analytic results; it can be seen equivalently from
numerical Monte Carlo measurements on rectangular $L_x \times L_y$ patches
after an extrapolation to $L_x=\infty$.  Indeed, Monte Carlo measurements would
be the standard method for this purpose since they are not limited, as the
exact analytic calculations are, to a rather small range of $L_y$ values.  
However, the value of discussing this with exact results is that the reader 
can verify the conclusions directly rather than having to reproduce them with 
another Monte Carlo study.

The organization of the paper is as follows.  In section 2 we discuss some
generalities of our approach.  In section 3 we prove that if one uses free
(periodic) boundary conditions in the $y$ direction transverse to the length of
the infinite strip, then $W$ approaches its 2D thermodynamic limit
monotonically (nonmonotonically).  Section 4 contains the numerical results for
strips of the square, triangular, and honeycomb lattices with free transverse
boundary conditions, while section 5 contains analogous results for periodic
transverse boundary conditions.  In section 6 we remark on how these strip
studies can be used to determine the central charge for cases with critical
ground states.  Our conclusions are presented in section 7.

\section{$W$ on Strip Graphs and the Approach to 2D Thermodynamic Limit}

The usual thermodynamic limit of the Potts antiferromagnet or other statistical
mechanical model on the 2D lattice $\Lambda$ involves taking $L_x \to \infty$
and $L_y \to \infty$ with fixed $L_y/L_x = \rho$, where ($\rho \ne 0, \infty$).
The question of how various thermodynamic quantities approach their 2D limits
as a function of $\rho$ has been of interest for many years (e.g., Ref. 
\cite{ff} for the Ising model).  As noted in the introduction, a different
way to approach the 2D thermodynamic limit is via a sequence of infinitely long
strips of progressively greater and greater widths.  That this is different is
clear from the fact that for each such strip, regardless of how large $L_y$ is,
the ratio $L_y/L_x=0$.  We picture the strip graphs as extending longitudinally
in the horizontal ($x$) direction and having a width of $L_y$ vertices in the
vertical direction.  {\it A priori}, it is not clear that this different
approach will yield results that are useful to the study of the 2D
thermodynamic limit, because, for a given thermodynamic quantity of interest,
these results might be dominated by the fundamentally 1D nature of the infinite
strip.  Indeed, to illustrate a case where it is not useful, consider a model,
such as a discrete ferromagnet, which has a second-order phase transition at
some critical temperature $T_c(\Lambda)$ on a 2D lattice $\Lambda$, and assume
that there is no exact solution of this model.  If one were to try to employ
exact solutions of the model on infinitely long strips of lattice type
$\Lambda$ to determine $T_c(\Lambda)$ for the 2D lattice, one would get the 1D
result $T_c=0$ for any finite value of $L_y$.  Hence, for $T_c(\Lambda)$ this
method would not give any useful information.  However, as we shall show, the
situation is very different with the ground state entropy $S_0(\Lambda)$; for 
this quantity, one can use results on infinite strips to get quite accurate 
values even for rather modest strip widths.

The existence of the thermodynamic limit for the 2D lattice $\Lambda$ means
that the maximal finite real $q$ where $W(\Lambda,q)$ is nonanalytic,
$q_c(\Lambda)$, is independent of the boundary conditions used in taking the 2D
thermodynamic limit \cite{w}.  Let us denote the $W$ function for the $L_x
\times L_y$ strip of the lattice of type $\Lambda$ as $W(\Lambda(L_x \times
L_y),BC_x,BC_y,q)$.  We observe here that for physical (positive integral) $q >
q_c(\Lambda)$, in the limit $L_x \to \infty$, this $W$ function is independent
of the boundary conditions used in the $x$ direction.  This is also true for
real $q > q_c(\Lambda)$ (and more generally, in the region of the complex $q$
plane denoted $R_1$ in our previous studies \cite{w}).  We thus introduce a
more compact notation for the $W$ function on infinitely long strips: \beq
W(\Lambda(L_y),BC_y,q) \equiv \lim_{L_x \to \infty} W(\Lambda(L_x \times
L_y),BC_x,BC_y,q)
\label{wfun}
\eeq 
Indeed, let $\Lambda_d$ be an infinite $d$-dimensional lattice and
$\Lambda_{d-1,L_d}$ be a slab of a $(d-1)$-dimensional lattice, infinite in
$d-1$ dimensions and of finite thickness $L_d$ in the $d$'th dimension.  For
physical $q > q_c(\Lambda_d)$, the value of $W(\Lambda_{d-1,L_d},q)$ is
independent of the boundary conditions used for the $(d-1)$ directions when
taking $L_j \to \infty$ for $1 \le j \le d-1$.  As we have discussed before
\cite{w}, this is not true for all $q \in {\mathbb C}$; however, here we deal
only with physical $q$ values.

\section{Issue of Monotonicity of Approach to 2D Thermodynamic Limit}

In this section we show that for free (periodic) boundary conditions in the 
$y$ direction, $W$ for infinitely long strips of width $L_y$ approaches its 2D
thermodynamic limit monotonically (nonmonotonically) as $L_y \to \infty$.

\subsection{Monotonic Approach for FBC$_{\lowercase{y}}$} 

We begin with the case of free boundary conditions and state the following
theorem:

\begin{flushleft}
Theorem 1
\end{flushleft}

Let $\Lambda_{d-1,L_d}$ denote a regular lattice graph of infinite extent in
$d-1$ dimensions and width (thickness) $L_d$ in the $d$'th dimension.  Let the
boundary conditions in the $d$'th direction be free and the boundary conditions
in each of the first $d-1$ be (separately) free or periodic.  (Note that
$\Lambda_{d-1,1} \equiv \Lambda_{d-1}$ and $\Lambda_{d-1,\infty}=\Lambda_{d}$.)
Then for fixed $q > q_c(\Lambda_d)$, $W(\Lambda_{d-1,L_d},q)$ is a 
monotonically decreasing function of $L_d$ for $1 \le L_d \le \infty$.

\vspace{3mm}

\begin{flushleft}
Proof
\end{flushleft}

   We shall prove the theorem for the case $d=2$; its generalization to $d \ge
3$ will be obvious.  We consider a finite strip graph of the lattice, of size
$L_x \times L_y$, where the longitudinal direction is $x$.  Assume that one has
made an allowed coloring of this graph.  Now connect another layer of sites to
the layer that formerly constituted the top layer of sites on the strip. The
coloring of this new layer of sites imposes additional constraints on the
coloring of the original strip, and excludes a certain subset of what were
previously allowed colorings.  Thus, the fraction of sites on the augmented
graph that have more constraints increases; i.e., the sites on the upper and
lower edges, which have fewer constraints on their coloring because of the free
transverse boundary conditions, constitute a progressively smaller fraction of
the total number of sites as $L_y$ increases.  Hence, the chromatic polynomial
per site, $P(\Lambda,[L_x \times L_y],q)^{1/n}$, decreases.  This inequality
holds for arbitrary $L_x$.  Taking the limit as $L_x \to \infty$ and using the
definition (\ref{w}), one obtains the theorem for the case $d=2$.  A
straightforward generalization of this argument proves the theorem for $d \ge
3$.  $\Box$

\vspace{3mm}

A corollary of this theorem is that if one compares $W$ on two infinite
lattices of the same type and of different dimensions, such as $d$-dimensional
cartesian lattices then, for fixed $q > q_c(\Lambda)$, 
\beq
W(\Lambda_d,q) < W(\Lambda_{d'},q) \quad {\rm if} \quad d > d'
\label{winf}
\eeq 
To prove this corollary, one starts with $d'=d-1$ and (i) constructs
$\Lambda_d$ from $\Lambda_{d-1}$ by imposing free boundary conditions in the
$d$'th direction and adding layers of vertices in this $d$'th direction, (ii)
uses the monotonicity relation of theorem 1 for the quantities
$W(\Lambda_{d-1,L_d},q)$, and (iii) takes the number of added layers in the
$d$'th direction to infinity to get $\Lambda_d$.  The monotonicity relation for
the infinite lattices (\ref{winf}) was previously noted by Chow
and Wu \cite{chowwu}. It is important to observe that the monotonicity relation
(\ref{winf}) does not imply our monotonicity theorem 1.  This is clear from the
fact that the inequality (\ref{winf}) holds independent of the boundary
conditions that one uses to define the respective thermodynamic limits on
$\Lambda_d$ and $\Lambda_{d-1}$, whereas, on the contrary, the inequality in
our theorem 1 does not apply if one uses periodic boundary conditions for the
$d$'th direction of the $(d-1)$-dimensional strip or slab of width $L_d$ (see
below).

\vspace{3mm} 

\subsection{Non-monotonic Approach for PBC$_{\lowercase{y}}$} 

\begin{figure}
\vspace{-2cm}
\centering
\leavevmode
\epsfxsize=3.0in
\begin{center}
\leavevmode
\epsffile{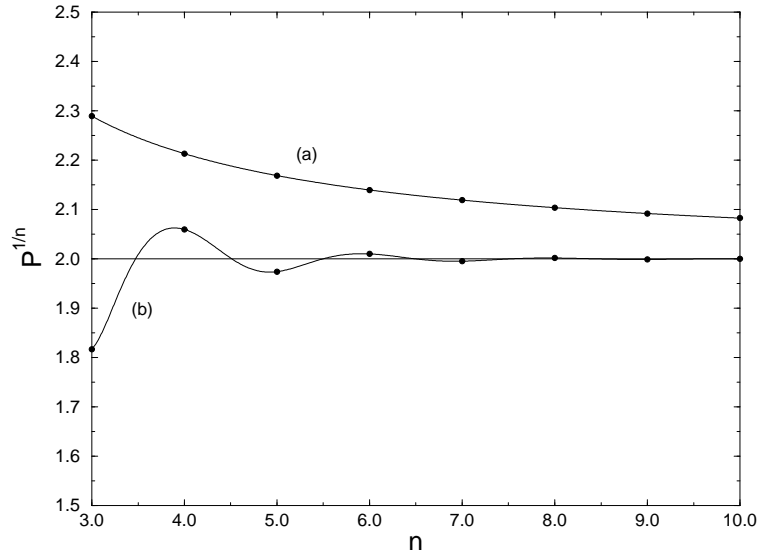}
\end{center}
\vspace{-2cm}
\caption{\footnotesize{Plots of (a) eq. (\ref{wtree}) and (b) 
eq. (\ref{wcircuit}) for $q=3$, as functions of real $n$, 
together with physical values for integer $n$.  Horizontal line is the 
asymptote, $W(\Lambda_1,q=3)=2$.}}
\label{wcnq3}
\end{figure}

Next, we show that a similar monotonicity result does not hold if one
imposes periodic boundary conditions in the $d$'th direction.  This is clear
from the proof, since the greater freedom in coloring the sites on the boundary
layer in the $d$'th dimension played a crucial role, but if one imposes
periodic boundary conditions in the $d$'th direction, there is no such boundary
layer.  The simplest illustration is provided by the case $d=1$, for which
\cite{w} $q_c(\Lambda_1)=2$ and 
\beq
W(\Lambda_1,q)=q-1
\label{wlam1}
\eeq
For free boundary conditions, the function that enters on the right-hand side 
of eq. (\ref{w}) is 
\beq
P((\Lambda_1)_n,FBC,q)^{1/n} = q^{1/n}(q-1)^{1-\frac{1}{n}}
\label{wtree}
\eeq
For fixed $q \ge q_c(\Lambda_1)$, this is a monotonically decreasing
function of $n$ as $n$ increases from 1 to infinity. However, if we impose
periodic boundary conditions, i.e. deal with an $n$-vertex circuit graph
$C_n$ \footnote{Parenthetically, we note that $C_n$ is only a (proper) graph
for $n \ge 3$ since the strict mathematical definition of a graph
forbids (i) any multiple bond connecting a given pair of vertices (present for
$C_{n=2}$) and (ii) any bond going out from a given vertex and looping back to
the same vertex (present for $C_{n=1}$).  This is not important for our
demonstration of non-monotonicity.}, then, the function that enters on the
right-hand side of eq. (\ref{w}) is 
\beq
P((\Lambda_1)_n,PBC,q)^{1/n} = (q-1)\biggl [ 
1 + (-1)^n(q-1)^{-(n-1)} \biggr ]^{1/n}
\label{wcircuit}
\eeq 
This is a non-monotonic function of $n$.  For example, for the lowest
value of $q$ where the 1D Potts AF has nonzero ground state entropy, viz.,
$q=3$, for which the $n \to \infty$ limit is $W(\Lambda_1,q=3)=2$,
eq. (\ref{wcircuit}) exhibits the non-monotonic behavior indicated by the
values $6^{1/3} = 1.817..$ for $n=3$, $(18)^{1/4} = 2.060..$ for $n=4$,
$(30)^{1/5} = 1.974..$ for $n=5$, $(66)^{1/6} = 2.010..$ for $n=6$,
etc. Similar non-monotonic behavior occurs for higher values of $q$.  Looking
at subsequences, we find that $P((\Lambda_1)_n,PBC,q)^{1/n}$ is a monotonically
increasing function of $n$ for odd $n \ge 3$ and a monotonically decreasing
function of $n$ for even $n \ge 2$.  This is connected with the fact that the
circuit graph $[(\Lambda_1)_n,PBC] = C_n$ with odd (even) $n$ has chromatic
number $\chi=3$ ($\chi=2$).  The different behaviors of
$P((\Lambda_1)_n,FBC,q)^{1/n}$ and 
$P((\Lambda_1)_n,PBC,q)^{1/n}$ can be seen in
a more general context by analytically continuing eqs. (\ref{wtree}) and
(\ref{wcircuit}) from $n \in {\mathbb Z}_+$ to $n \in {\mathbb R}_+$ and
plotting them as functions of $n$ (in the second case, since $P(C_n,q)=
(q-1)^n+(-1)^n(q-1)$ is complex for $n \notin {\mathbb Z}$, we plot
$|P(C_n,q)|^{1/n}$).  This is shown in Fig. \ref{wcnq3}.  One notices that
although eq. (\ref{wcircuit}) for periodic boundary conditions behaves
non-monotonically, it approaches the $n=\infty$ value $W(\Lambda_1,q=3)=2$
considerably more rapidly than the FBC expression, eq. (\ref{wtree}).
As one increases $q$ beyond 3, the first maximum in $|P(C_n,q)|^{1/n}$ moves
slightly leftward, and the oscillations damp out faster.  As we shall show in
the tables below, a similar difference holds between the behavior of
$W(\Lambda(L_y),BC_y,q)$ for FBC$_y$ and PBC$_y$.

\section{Quantitative Results for Strips with FBC$_{\lowercase{y}}$}

   One would like to go beyond the general inequality in Theorem 1 to obtain an
explicit numerical determination of the dependence of $W$ on $L_d$.  We do this
here for $d=2$ and, in particular, for the square (sq), triangular (t), and
honeycomb (hc) lattices.  For the strip graph of each type $G_s$, we define the
ratio
\beq 
R_W(\Lambda(L_y),BC_y,q) =
\frac{W(\Lambda(L_y),BC_y,q)}{W(\Lambda,q)}
\label{rw}
\eeq 
In the appendix we list the exact analytic expressions for
$W(\Lambda(L_y),FBC_y,q)$ for $\Lambda=sq,t,hc$ and the $L_y$ values used here.
In Table \ref{sqtable} we show a numerical comparison for strips of the square
lattice (along the row direction) for $1 \le L_y \le 4$ and $3 \le q \le 10$.
The exact value $W(sq,q=3)=(4/3)^{3/2}=1.53960...$ is from Ref. \cite{lieb},
while the values of $W(sq,q)$ for $4 \le q \le 10$ are from our Monte Carlo
measurements in Ref. \cite{w}.  Using the conservatively quoted uncertainties
that we gave for the Monte Carlo measurements, it would follow that the
corresponding uncertainties in the ratios (\ref{rw}) are $\sim (3-4) \times
10^{-4}$; with less conservative estimates of uncertainties in the Monte Carlo
measurements, the resultant uncertainties in these ratios would be smaller.  In
Tables \ref{tritable} and \ref{hctable} we give the analogous comparisons for
strips of the triangular lattice of widths $L_y=2,3,4$ and of the honeycomb
lattice for widths $L_y=2$ and 3.  In all of these cases, one observes that,
for fixed $q$, the agreement with the infinite-lattice value gets better as the
width increases and for fixed width, the agreement gets better as the value of
$q$ increases.  These comparisons show that the approach to the 2D
thermodynamic limit is reasonably rapid even for free boundary conditions in
the transverse direction. For example, for an $\infty \times 4$ strip of the
square lattice for $3 \le q \le 5$, the $W$ values are within about 5 \% of
their respective values for the infinite 2D lattice.

\pagebreak

\begin{table}
\caption{\footnotesize{Comparison of values of $W(sq(L_y),FBC_y,q)$ with
$W(sq,q)$ for $3 \le q \le 10$.  For each value of $q$, the quantities in the
upper line are identified at the top and the quantities in the lower line are
the values of $R_W(sq(L_y),FBC_y,q)$. The FBC$_y$ is symbolized as $F$ in the
table.}}
\begin{center}
\begin{tabular}{cccccccc}
$q$ & $W(sq(1),F,q)$ & $W(sq(2),F,q)$ & $W(sq(3),F,q)$ & $W(sq(4),F,q)$ &  
$W(sq,q)$ & $W(sq,q)_s$ & $W(sq,q)_\ell$ \\
3 &   2    & 1.73205   & 1.65846   & 1.624945 & 1.53960..  & 1.53960.. &
  1.50000 \\
  & 1.299  & 1.125    & 1.077   & 1.055       & 1          & 1       \\
4 &   3    & 2.64575   & 2.53800   & 2.48590   & 2.3370(7)  & 2.3361 &
  2.33333 \\
  & 1.284  & 1.132     & 1.086     & 1.064     & 1          & - & -   \\
5 &   4    & 3.60555   & 3.48304   & 3.42336   & 3.2510(10) & 3.2504 &
  3.25000 \\
  & 1.230  & 1.109     & 1.071     & 1.053     & 1          & - & -   \\
6 &   5    & 4.58258   & 4.45136   & 4.38717   & 4.2003(12) & 4.2001 &
  4.20000 \\
  & 1.190  & 1.091     & 1.060     & 1.0445     & 1          & - & -   \\
7 &   6    & 5.56776   & 5.43073   & 5.36348    & 5.1669(15) & 5.1667 &
  5.16667 \\
  & 1.161  & 1.078     & 1.051     & 1.038     & 1          & - & -   \\
8 &   7    & 6.55744   & 6.41623   & 6.34677   & 6.1431(20) & 6.1429 &
  6.14286 \\
  & 1.1395  & 1.067    & 1.0445     & 1.033     & 1          & - & -   \\
9 &   8    & 7.54983   & 7.40548    & 7.33434   & 7.1254(22) & 7.1250 &
  7.12500 \\
  & 1.123  & 1.060     & 1.039     & 1.029     & 1          & - & -   \\
10 &   9   & 8.54400   & 8.39720   & 8.324745  & 8.1122(25) & 8.1111 &
   8.11111 \\
  & 1.109  & 1.053     & 1.035     & 1.026     & 1          & - & -   \\
\hline
\end{tabular}
\end{center}
\label{sqtable}
\end{table}

\pagebreak

\begin{table}
\caption{\footnotesize{Comparison of values of $W(t(L_y),FBC_y,q)$ with 
$W(t,q)$ for $3 \le q \le 10$.  For each value of $q$, 
the quantities in the
upper line are identified at the top and the quantities in the lower line are
the values of $R_W(t(L_y),FBC_y,q)$. The FBC$_y$ is symbolized as $F$ in the
table.}}
\begin{center}
\begin{tabular}{cccccc}
$q$ & $W(t(2),F,q)$ & $W(t(3),F,q)$ & $W(t(4),F,q)$ & $W(t,q)$ & 
$W(t,q)_\ell$ \\
4  &   2       & 1.77173   & 1.67619   & 1.46100    &  1.333333 \\
   & 1.369     & 1.213     & 1.147     & 1          &  -        \\
5  &   3       & 2.72998   & 2.60495   & 2.26411    &  2.250000 \\
   & 1.325     & 1.206     & 1.151     & 1          &  -        \\
6  &   4       & 3.71457   & 3.579715  & 3.20388    &  3.200000 \\
   & 1.248     & 1.159     & 1.117     & 1          &  -        \\
7  &   5       & 4.70571   & 4.56515   & 4.16819    &  4.166667 \\
   & 1.200     & 1.129     & 1.095     & 1          &  -        \\
8  &   6       & 5.69974   & 5.55530   & 5.14358    &  5.142857 \\
   & 1.167     & 1.108     & 1.080     & 1          &  -        \\
9  &   7       & 6.695395  & 6.54810   & 6.12539    &  6.125000 \\
   & 1.143     & 1.093     & 1.069     & 1          &  -        \\
10 &   8       & 7.69208   & 7.54259   & 7.11134    &  7.111111 \\
   & 1.125     & 1.082     & 1.061     & 1          &  -        \\
\hline
\end{tabular}
\end{center}
\label{tritable}
\end{table}

\pagebreak

\begin{table}
\caption{\footnotesize{Comparison of values of $W(hc(L_y),FBC_y,q)$ with 
$W(hc,q)$ for $3 \le q \le 10$.  For each value of $q$,
the quantities in the
upper line are identified at the top and the quantities in the lower line are
the values of $R_W(hc(L_y),FBC_y,q)$. The FBC$_y$ is symbolized as $F$ in the
table.}}
\begin{center}
\begin{tabular}{cccccc}
$q$ & $W(hc(2),F,q)$ & $W(hc(3),F,q)$ & $W(hc,q)$ &
$W(hc,q)_{ser.}$ & $W(hc,q)_\ell$ \\
3  & 1.82116   & 1.76567   & 1.6600(5)  & 1.6600  & 1.658312  \\
   & 1.097     & 1.064     & 1          & -       & -         \\
4  & 2.79468   & 2.72942   & 2.6038(7)  & 2.6034  & 2.603417  \\
   & 1.073     & 1.048     & 1          & -       & -         \\
5  & 3.78389   & 3.71448   & 3.5796(10) & 3.5795  & 3.579455  \\
   & 1.057     & 1.038     & 1          & -       & -         \\
6  & 4.77760   & 4.70568   & 4.5654(15) & 4.5651  & 4.565085  \\
   & 1.046     & 1.031     & 1          & -       & -         \\
7  & 5.77336   & 5.69973   & 5.5556(17) & 5.5553  & 5.555278  \\
   & 1.039     & 1.026     & 1          & -       & -         \\
8  & 6.77028   & 6.69539   & 6.5479(20) & 6.5481  & 6.548095  \\
   & 1.034     & 1.023     & 1          & -       & -         \\
9  & 7.76793   & 7.69208   & 7.5424(22) & 7.5426  & 7.542587  \\
   & 1.030     & 1.020     & 1          & -       & -         \\
10 & 8.76607   & 8.68945   & 8.5386(25) & 8.5382  & 8.538222  \\
   & 1.027     & 1.018     & 1          & -       & -         \\
\hline
\end{tabular}
\end{center}
\label{hctable}
\end{table}

\section{Quantitative Results for Strips with PBC$_{\lowercase{y}}$}

In Tables \ref{sqtablepbc} and \ref{tritablepbc} we present similar results for
infinite strips with periodic boundary conditions in the transverse ($y$)
direction.  The exact analytic expressions that we use for these tables are
given in the Appendix.  As is mentioned in the Appendix, for a strip of the
square lattice with PBC$_y$ and cross sections forming triangles, depending on
one's labelling conventions, this corresponds to $L_y=3$ or $L_y=4$, where in
the latter case, one interprets the periodic boundary conditions as identifying
the top and bottom vertices for each value of $x$.  A similar comment applies
for a strip with PBC$_y$ and transverse cross sections forming squares.  For
the table, we use the convention of choosing the smaller of the respective
values of $L_y$.  We find that for a given $q$, $W$ approaches its 2D value
$W(\Lambda,q)$ much more rapidly with periodic rather than free transverse
boundary conditions: for the modest width $L_y=4$, $W$ is within ${\cal
O}(10^{-3})$ of its 2D value for moderate $q$.  The finding that the periodic
boundary conditions in the transverse direction yield a more rapid approach to
the 2D thermodynamic limit than the free boundary conditions is not, in itself,
a surprise; this is in accord with a wealth of past experience with statistical
mechanical models on finite-size lattices.  What is remarkable is how rapid in
absolute terms this approach is.  Of course, one can also consider larger
values of $L_y$, but the strikingly rapid approach to the 2D thermodynamic
limit is already fully demonstrated by the range of $L_y$ that we have
considered.

\begin{table}
\caption{\footnotesize{Comparison of values of $W(sq(L_y),PBC_y,q)$ with
$W(sq,q)$ for $3 \le q \le 10$.  For each value of $q$, the quantities in the
upper line are identified at the top and the quantities in the lower line are
the values of $R_W(sq(L_y),PBC_y,q)$. The PBC$_y$ is symbolized as $P$ in the
table.}}
\begin{center}
\begin{tabular}{cccc}
$q$ & $W(sq(3),P,q)$ & $W(sq(4),P,q)$ & $W(sq,P,q)$ \\
3 &   1.25992  & 1.58882   & 1.53960..  \\
  &   0.8183   & 1.032     & 1          \\
4 &   2.22398  & 2.37276   & 2.3370(7)  \\
  &   0.9516   & 1.015     & 1          \\
5 &   3.17480  & 3.26878   & 3.2510(10) \\
  &   0.9766   & 1.0055    & 1          \\
6 &   4.14082  & 4.21082   & 4.2003(12) \\
  &   0.9858   & 1.002505  & 1          \\
7 &   5.11723  & 5.17377   & 5.1669(15) \\
  &   0.9904   & 1.0013    & 1          \\
8 &   6.10017  & 6.14792   & 6.1431(20) \\
  &   0.9930   & 1.0008    & 1          \\
9 &   7.08734  & 7.12881   & 7.1254(22) \\
  &   0.9947   & 1.0005    & 1          \\
10 &  8.07737  & 8.11409   & 8.1122(25) \\
  &   0.9957   & 1.0002    & 1          \\
\hline
\end{tabular}
\end{center}
\label{sqtablepbc}
\end{table}

\pagebreak

\begin{table}
\caption{\footnotesize{Comparison of values of $W(t(L_y),PBC_y,q)$ with
$W(t,q)$ for $4 \le q \le 10$.   For each value of $q$, the quantities in the
upper line are identified at the top and the quantities in the lower line are
the values of $R_W(t(L_y),PBC_y,q)$.  The PBC$_y$ is symbolized as $P$ in the
table.}}
\begin{center}
\begin{tabular}{cccc}
$q$ & $W(t(3),P,q)$ & $W(t(4),P,q)$ & $W(t,q)$ \\
4  & 1.58740   & 1.18921    & 1.46100     \\
   & 1.0865    & 0.8140     & 1           \\
5  & 2.35133   & 2.21336    & 2.26411     \\
   & 1.0385    & 0.9776     & 1           \\
6  & 3.23961   & 3.185055   & 3.20388     \\
   & 1.0112    & 0.9941     & 1           \\
7  & 4.17934   & 4.15965    & 4.16819     \\
   & 1.0027    & 0.99795    & 1           \\
8  & 5.14256   & 5.13936    & 5.14358     \\
   & 0.99980   & 0.9992     & 1           \\
9  & 6.11803   & 6.12324    & 6.12539     \\
   & 0.99880   & 0.99965    & 1           \\
10 & 7.10059   & 7.11027    & 7.11134     \\
   & 0.99849   & 0.99985    & 1           \\
\hline
\end{tabular}
\end{center}
\label{tritablepbc}
\end{table}

\section{Cases with Critical Ground States}

For certain 2D lattices $\Lambda$ and values of $q$, the $q$-state Potts 
antiferromagnet has a critical ground state, i.e., as $T \to 0$, a correlation
length $\xi$ defined, say, by a spin-spin correlation function, goes to 
infinity.  Normally, in statistical mechanics, for a given dimensionality $d$
and symmetry group $G$, second-order phase transitions can be described by a
universality class representing a fixed point of the renormalization group.
Conformal field theory methods have provided a powerful way to understand these
universality classes and the associated critical exponents in terms of Virasoro
algebras with given central charges and scaling dimensions \cite{fqs}. In
addition to phase transitions involving ferromagnetic long range order at low
temperatures, this is also true of antiferromagnetic transitions on bipartite
lattices, but the situation is more complicated on nonbipartite lattices, as is
illustrated by the fact that the isotropic Ising antiferromagnet on the
triangular lattice has no finite-temperature phase transition but is critical
at $T=0$.  

   The $q=3$ Potts antiferromagnet on the square lattice has a critical
ground state with central charge $c=1$, as a consequence of the fact that at 
$T=0$ this model can be mapped to a critical six-vertex model \cite{lieb}. 
From the exact solution in Ref. \cite{baxter}, it can be argued that 
the $q=4$ Potts antiferromagnet on the triangular lattice is also critical, 
which is closely related to the fact that the $q=3$ Potts antiferromagnet on 
the kagom\'e lattice is critical at $T=0$ \cite{henley}.  
We recall that given the Virasoro algebra with central extension
\beq
[L_m, L_n] = (m-n)L_{m+n} + \frac{c}{12}m(m^2-1)\delta_{m+n,0}
\label{virasoro}
\eeq
and the corresponding Kac-Moody algebra realized at level $k$
\beq
[J_m^a, J_n^b] = c^{abc}J_{m+n}^c + \frac{1}{2}kn \delta^{ab}\delta_{m+n,0}
\label{kacmoody}
\eeq
with structure constants $c_{abc}$, as connected via the Sugawara relation 
(e.g., \cite{fqs}) 
\beq
L_n = -\frac{1}{C_2(g) + k}\sum_{m=-\infty}^{\infty} : J_m^a J_{n-m}^a :
\label{sugawara}
\eeq
it follows that 
\beq
c = \frac{dim(g)}{C_2(g)/k + 1}
\label{crelation}
\eeq
where $C_2(g)$ is the quadratic Casimir operator for the algebra $g$.
In particular, 
\beq
g=su(M)_{k=1} \ \ \Longrightarrow  \quad c=M-1
\label{gcm}
\eeq
Hence, from eq. (\ref{gcm}) together with the finding \cite{henley} that the 
Kac-Moody algebra is $su(3)_{k=1}$ for the $T=0$ $q=3$ Potts AF on the 
kagom\'e lattice \cite{henley}, it follows that $c=2$ for this critical ground
state.  Given that there is a close connection between the Potts
antiferromagnets with $q=3$ on the kagom\'e lattice and with $q=4$ on the
triangular lattice, which leads to the relation $W(kag,q=3)=W(tri,q=4)^{1/3}$
\cite{baxter}, this suggests that this value of $c=2$ also holds for the
$T=0$, $q=4$ Potts AF on the triangular lattice. 

Here we would like to point out that determinations of the ground state entropy
on infinitely long strips of finite width can be used to obtain the central
charge $c$ for Potts antiferromagnets with critical ground states.  If one
considers the model on a lattice of size $L_x \times L_y$, then, in the limit 
as $L_x \to \infty$, one has \cite{cardy}
\beq
f_{strip,L_y} = f_{bulk} + \frac{f_{surf.}}{L_y} + \frac{\Delta}{L_y^2} +
O(L_y^{-3})
\label{frel}
\eeq
where $f_{surf.}=0$ is nonzero (zero) for free (periodic) boundary
conditions in the $y$ direction and
\beq
\Delta = \left \{ \begin{array}{ll}
      \frac{\pi}{6}c & \mbox{for PBC$_y$} \\
     \frac{\pi}{24}c & \mbox{for FBC$_y$}
     \end{array} \right .
\label{delta}
\eeq
For the critical ground states of interest here, viz., 
$q=3,4,3$ on the square, triangular, and kagom\'e lattices, respectively, as
well as other possible 2D cases, the Potts antiferromagnet exhibits ground 
state entropy without frustration, and the reduced free energy (per site)
$f = \lim_{N \to \infty} N^{-1} \ln Z$ is given simply by the ground
state entropy: $f(\Lambda,q)_{PAF}  = S_0(\Lambda,q)_{PAF}/k_B$, 
Hence, eq. (\ref{frel}) becomes
\beq
S_{strip,L_y} = S_{bulk} + \frac{S_{surf.}}{L_y} + \frac{\Delta}{L_y^2} +
O(L_y^{-3})
\label{srel}
\eeq
Thus, calculations of $S_{strip,L_y}$ for several different values of $L_y$ can
yield $c$.  Normally, one would do this via the most general and robust method,
namely Monte Carlo simulations.  For small $L_y$ values, we have shown that it
is actually possible to get exact analytic results, but this method is not 
competitive with Monte Carlo simulations for strips with larger values
of $L_y$.  One might note in passing that for the $q=3$ Potts AF on the 
infinite square strip with PBC$_y$ and cross sections forming squares, taking 
$L_y=4$, using the fact that $S_{surf.}=0$ in this case, and dropping terms of
order $L_y^{-3}$ in eq. (\ref{srel}), we obtain the estimate $c=0.96$, quite
close to the exact value $c=1$.

\section{Conclusions} 

In summary, we have studied a different type of approach to the 2D
thermodynamic limit for the ground state entropy, or equivalently, the ground
state degeneracy per site, $W$, of the $q$-state Potts antiferromagnet, using
infinitely long strips of increasing widths.  We have found that the approach
of $W$ to its 2D thermodynamic limit is quite rapid; for moderate values of $q$
and widths $L_y \simeq 4$, $W(\Lambda_{L_y},q)$ is within about 5 \% and
${\cal O}(10^{-3})$ of the 2D value for free and periodic boundary conditions,
respectively.  We have also proved that the approach of $W$ to the 2D
thermodynamic limit is monotonic (non-monotonic) for free (periodic) boundary
conditions in the transverse direction.  Finally, we have noted that these
ground state entropy determinations on infinite strips can be used to obtain
the central charge for cases with critical ground states.

\vspace{10mm}

We are grateful to Prof. M. Ro\v{c}ek for the collaborative work on Ref.
\cite{strip}.  This research was supported in part by the NSF grant
PHY-97-22101.

\vspace{6mm}

\section{Appendix}

We gather together here the exact analytic formulas on which our numerical
tables are based.  It should be emphasized that the entries in these tables and
the resultant conclusions about the rapidity of the approach of $W$ to the 2D
thermodynamic limit for infinitely long strips with free or periodic transverse
boundary conditions could also have been obtained using purely numerical Monte
Carlo calculations.  The usefulness of the analytic formulas (which are
elementary for $L_y=1,2$) is just that they enable one to check the results
more directly.

\subsection{Square Lattice, FBC$_{\lowercase{y}}$}

For infinitely long strips of the square lattice with FBC$_y$, we have 
\beq
W(sq(L_y=1),FBC_y,q) = q-1
\label{wsqly1}
\eeq
\beq
W(sq(L_y=2),FBC_y,q) = (q^2-3q+3)^{1/2}
\label{wsqly2}
\eeq
\beqs
W(& & sq(L_y=3),FBC_y,q) = 
2^{-1/3}\Biggl [ (q-2)(q^2-3q+5) \cr\cr
& & + \Bigl [(q^2-5q+7)(q^4-5q^3+11q^2-12q+8) \Bigr ]^{1/2} \Biggr ]^{1/3}
\label{wsqly3fbc}
\eeqs
$W(sq(L_y=4),FBC_y,q)$ is given by the maximal root of the cubic equation
\beq
\xi^3+b_{sq(4),1}\xi^2 + b_{sq(4),2}\xi + b_{sq(4),3} = 0
\label{xieqsqly4}
\eeq
where the coefficients $b_{sq(4),k}$, $k=1,2,3$ were listed in 
Ref. \cite{strip}.

\subsection{Triangular Lattice, FBC$_{\lowercase{y}}$ }

For the triangular lattice strips with FBC$_y$, we have 
\beq
W(t(L_y=2),FBC_y,q) = q-2
\label{wtrily2}
\eeq
\beqs
W(& & t(L_y=3),FBC_y,q) = 
2^{-1/3}\Biggl [ (q^3-7q^2+18q-17) \cr\cr
& & + \Bigl [q^6-14q^5+81q^4-250q^3+442q^2-436q+193 \Bigr ]^{1/2} 
\Biggr ]^{1/3}
\label{wtrily3}
\eeqs
$W(t(L_y=4),FBC_y],q)$ is given by the maximal root of the quartic equation
\beq
\xi^4+b_{t(4),1}\xi^3 + b_{t(4),2}\xi^2 + b_{t(4),3}\xi + b_{t(4),4} = 0
\label{xieqtrily4}
\eeq
where the $b_{t(4),k}$, $k=1,..,4$ were listed in Ref. \cite{strip}. 

\subsection{Honeycomb Lattice, FBC$_{\lowercase{y}}$ }

For the honeycomb lattice strips with FBC$_y$, we have 
\beq
W(hc(L_y=2),FBC_y,q) = (q^4-5q^3+10q^2-10q+5)^{1/4}
\label{whcly2}
\eeq
$W(hc(L_y=3),FBC_y,q)$ is given by the maximal root of the cubic equation
\beq
\xi^3+b_{hc(3),1}\xi^2 + b_{hc(3),2}\xi + b_{hc(3),3} = 0
\label{xieqhcly3}
\eeq
where the $b_{hc(3),k}$, $k=1,2,3$ were listed in \cite{strip}

\subsection{Square Lattice, PBC$_{\lowercase{y}}$}

We first consider a strip of the square lattice with PBC$_y$ and transverse 
cross sections forming triangles.  Depending on one's labelling conventions, 
this corresponds to $L_y=3$ or $L_y=4$, where in the latter case, one 
interprets the periodic boundary conditions as identifying the top and 
bottom vertices for each value of $x$.  We calculate 
\beq
W(sq(L_y=3),PBC_y,q) = (q^3-6q^2+14q-13)^{1/3}
\label{wsqly3pbc}
\eeq
For the next larger size, i.e. transverse cross sections forming squares,
corresponding to $L_y=4$ or $L_y=5$ in the respective labelling conventions
described above, the $W$ function is given by \cite{strip} 
\beqs
W(& &sq(L_y=4),PBC_y,q = 2^{-1/4}\Biggl [ (q^4-8q^3+29q^2-55q+46)  \cr\cr
& & + \Bigl [q^8-16q^7+118q^6-526q^5+1569q^4-3250q^3+4617q^2-4136q+1776 
\Bigr ]^{1/2} \Biggr ]^{1/4}
\label{wsqly4pbc}
\eeqs

\subsection{Triangular Lattice, PBC$_{\lowercase{y}}$}

We next consider a strip of the triangular lattice with PBC$_y$, represented as
a square lattice with additional diagonal bonds from, say, the upper left to
lower right vertices of each square. 
 For the case where the transverse cross sections form 
triangles, corresponding to $L_y=3$ or $L_y=4$ in the above labelling
conventions, we calculate 
\beq
W(t(L_y=3),PBC_y,q) = (q^3-9q^2+29q-32)^{1/3}
\label{wtrily3pbc}
\eeq
For the next larger size, with transverse cross sections forming squares, $W$
is \cite{strip} 
\beqs
W(& & t(L_y=4),FBC_y],q) = 
2^{-1/4}(q-3)^{1/4}\Biggl [ (q^3-9q^2+33q-48)  \cr\cr
& & + (q-4)\Bigl [q^4-10q^3+43q^2-106q+129 \Bigr ]^{1/2} \Biggr ]^{1/4}
\label{wtrily4pbc}
\eeqs

\pagebreak

\vfill
\eject


\begin{thebibliography}{99}

\bibitem{lp}{L. Pauling, {\it The Nature of the Chemical Bond}
(Cornell Univ. Press, Ithaca, 1960), p. 466.}

\bibitem{liebwu}{E. H. Lieb and F. Y. Wu, in C. Domb and M. S. Green,
eds., {\it Phase Transitions and Critical Phenomena} (Academic Press,
New York, 1972) v. 1, p. 331.}

\bibitem{al}{M. Aizenman and E. H. Lieb, J. Stat. Phys. {\bf 24}, 279 (1981).}

\bibitem{chowwu}{Y. Chow and F. Y. Wu, Phys. Rev. {\bf B36}, 285 (1987);
Y. Chow, Discrete Math, {\bf 66} (1987) 51-58.}

\bibitem{potts}{R. B. Potts, Proc. Camb. Phil. Soc. {\bf 48} (1952) 106.}

\bibitem{wurev}{F. Y. Wu, Rev. Mod. Phys. {\bf 54}, 235 (1982);  errata,
{\it ibid}. {\bf 55}, 315 (1983).}

\bibitem{rtrev}{R. C. Read, J. Combin. Theory {\bf 4}, 52 (1968);
R. C. Read and W. T. Tutte, ``Chromatic Polynomials'',
in {\it Selected Topics in Graph Theory, 3}, eds. L. W. Beineke and
R. J. Wilson (Academic Press, New York, 1988.)}

\bibitem{lieb}{E. H. Lieb, Phys. Rev. {\bf 162}, 162 (1967).}

\bibitem{baxter}{R. J. Baxter, J. Phys. A {\bf 20}, 5241 (1987).}

\bibitem{series}{J. F. Nagle, J. Combin. Theory {\bf 10} (1971) 42;
G. A. Baker, Jr., J. Combin. Theory {\bf 10} (1971) 217.} 

\bibitem{kewser}{D. Kim and I. G. Enting, J. Combin. Theory, B {\bf 26},
327 (1979).}

\bibitem{ww}{R. Shrock and S.-H. Tsai,  Phys. Rev. {\bf E55}, 6791 (1997).}

\bibitem{w3}{R. Shrock and S.-H. Tsai, Phys. Rev. {\bf E56}, 2733 (1997).}

\bibitem{wn}{R. Shrock and S.-H. Tsai, Phys. Rev. {\bf E56},, 4111 (1997);
S.-H. Tsai, Phys. Rev. {\bf E57}, 2686 (1998).}

\bibitem{biggs}{N. L. Biggs, Bull. London Math. Soc. {\bf 9}, 54 (1977).}

\bibitem{cp}{X. Chen and C. Y. Pan, Int. J. Mod. Phys. {\bf B1}, 111
(1987); C. Y. Pan and X. Chen, {\it ibid.} {\bf B2}, 1503 (1988).}

\bibitem{wsk}{J.-S. Wang, R. H. Swendsen, and R. Koteck\'y,
Phys. Rev. B {\bf 42}, 2465 (1990).}

\bibitem{p3afhc}{R. Shrock and S.-H. Tsai, J. Phys. A {\bf 30}, 495 (1997).}

\bibitem{bds}{N. L. Biggs, R. M. Damerell, and D. A. Sands, J. Combin. Theory
B {\bf 12}, 123 (1972).}

\bibitem{bkw}{S. Beraha, J. Kahane, and N. Weiss, J. Combin. Theory B
{\bf 28}, 52 (1980).}

\bibitem{read91}{R. C. Read and G. F. Royle, in {\it Graph Theory,
Combinatorics, and Applications} (Wiley, New York, 1991), vol. 2, p. 1009.}

\bibitem{w}{R. Shrock and S.-H. Tsai, Phys. Rev. {\bf E55}, 5165 (1997).} 

\bibitem{wc}{R. Shrock and S.-H. Tsai, Phys. Rev. {\bf E56}, 1342 (1997).}

\bibitem{wa}{R. Shrock and S.-H. Tsai, Phys. Rev. {\bf E56}, 3935 (1997); 
J. Phys. A, in press; ITP-SB-98-17.}

\bibitem{strip}{M. Ro\v{c}ek, R. Shrock, and S.-H. Tsai, Physica {\bf A252},
505 (1998); Physica {\bf A259}, 367 (1998).}

\bibitem{hs}{R. Shrock and S.-H. Tsai, Physica {\bf A259}, 315 (1998).}

\bibitem{ff}{A. E. Ferdinand and M. E. Fisher, Phys. Rev. {\bf 185}, 832
(1969).} 

\bibitem{fqs}{A. A. Belavin, A. M. Polyakov, and A. B. Migdal, Nucl. Phys. 
{\bf B241}, 333 (1984); D. Friedan, Z. Ziu, and S. Shenker, Phys. Rev. Lett. 
{\bf 52}, 1575 (1984); C. Itzykson, H. Saleur, and J.-B. Zuber, 
{\it Conformal Invariance and Applications to Statistical Mechanics} 
(World Scientific, Singapore, 1988); J. Cardy, in C. Domb and J. L. Lebowitz,
eds., {\it Phase Transitions and Critical Phenomena} (Academic Press, New York,
1987), vol. 11, p. 55.} 

\bibitem{henley}{J. Kondev and C. Henley, Nucl. Phys. {\bf B464}, 540 (1996).}

\bibitem{parkwidom}{H. Park and M. den Nijs, Phys. Rev. {\bf B38}, 565 (1988);
J. Phys. A {\bf 22}, 3663 (1989);
H. Park and M. Widom, Phys. Rev. Lett. {\bf 63}, 1193 (1989)}

\bibitem{cardy}{J. L. Cardy, J. Phys. A {\bf 17}, L385, L961 (1984);
H. W. J. Bl\"ote, and M. P. Nightingale, Phys. Rev. Lett. {\bf 56}, 742 (1986);
I. Affleck, Phys. Rev. Lett. {\bf 56}, 746 (1986).}

\end{thebibliography}
\end{document}